\documentclass[twocolumn,pra]{revtex4}
\usepackage{amsmath}
\usepackage{amstext}
\usepackage{latexsym}
\usepackage{graphicx}
\usepackage{amsfonts}

\begin{document}
  
\title{
Quantum superpositions and entanglement of thermal states at high temperatures
and their applications to quantum information processing}
  
\author{Hyunseok Jeong and Timothy C. Ralph}
  
\affiliation{Centre for Quantum Computer Technology,
Department of Physics, University of Queensland, St Lucia, Qld 4072,
   Australia }
  
  \date{\today}  

\begin{abstract}
We study characteristics of 
superpositions and entanglement of
thermal states at high temperatures
and discuss their applications to quantum information processing.
We introduce thermal-state qubits and thermal-Bell states,
which are a generalization of pure-state
qubits and Bell states to thermal mixtures.
A scheme is then presented to discriminate between the four thermal-Bell states
without photon number resolving detection but with Kerr nonlinear
interactions and two single-photon detectors.
This enables one to perform quantum teleportation and gate operations 
for quantum computation with thermal-state qubits.
\end{abstract}

\maketitle

\section{Introduction}

In many problems considered within the framework of quantum physics,
physical systems are 
treated as pure states
that can be represented by state vectors, or equivalently, by wave functions.
Even though such an approach is simple and useful to address 
certain problems, it could often be quite different from real
conditions of physical systems.
This may be particularly true when one deals with macroscopic physical systems
in terms of quantum physics.
A macroscopic object is a complex open system which cannot avoid
continuous interactions with the environment.
Such a physical system is generally in a significantly mixed state
and cannot be represented by a state vector. 
In general, mixed states are subtle objects whose properties
are significantly more difficult to characterize than pure states.

Schr\"odinger's famous cat paradox
is a typical example
where a massive
classical object was assumed to be a pure state.
It describes a counter-intuitive feature of quantum physics
which dramatically appears when the
 principle of  quantum superposition is applied to macroscopic objects.
In the original paradox and its various explanations,
the initial cat isolated in the steel chamber is
considered a pure state that can be represented by 
a state vector such as $|alive\rangle$ (or a 
wave function such as $\psi_{alive}$).
The cat isolated from the environment is then assumed to
interact with a microscopic superposition state,
$(|g\rangle+|e\rangle)/\sqrt{2}$,
where $|g\rangle$ and $|e\rangle$ are the
ground and excited states of a two-level atom. 
The cat will be dead if the atom is found in the excited state,
$|e\rangle$, while it will remain alive if otherwise.
Thus in Schr\"odinger's
gedanken experiment the cat is 
entangled with the atom as   
$(|g\rangle|alive\rangle+|e\rangle|dead\rangle)\sqrt{2}$,
where the alive and dead statuses of the cat are described by
the state vectors $|alive\rangle$ and $|dead\rangle$. 
If one measures out the atomic system on the superposed basis,
$(|g\rangle\pm|e\rangle)/\sqrt{2}$,
the cat will be in a superposition of alive and dead states such as  
$(|alive\rangle\pm|dead\rangle)/\sqrt{2}$.
It is often argued that such superposed states and entangled
states can theoretically exist but 
are virtually impossible to observe
because one cannot perfectly isolate
a macroscopic object such as the cat from its environment
\cite{nielsen}.

However, this explanation is not fully satisfactory because 
the cat, a macroscopic object,
 is a complex open system which cannot be represented by a state vector.
One may argue that the cat could be assumed to be in
an {\it unknown} pure state such that
the cat was certainly alive but the exact state of the cat was unknown.   
However, the interactions between the cat and its environment
can cause the cat to become entangled with the environment \cite{Wiseman}.
In such a case, even though one can perfectly isolate the cat in the steel chamber from the 
enviroment,
the cat  will remain entangled with the environment due to
its pre-interactions with the environment.
Therefore, strictly speaking, even to assume a cat as an unknown pure state in the steel chamber is 
 not legitimate.
 Thus a key point here is that it is
unsatisfactory to describe
 the cat by a pure state such as
 $|alive\rangle$ and $|dead\rangle$.
We may need a more realistic assumption
that the ``cat'' in Schr\"odinger's paradox was in a
significantly mixed classical state.
An intriguing question is then whether the quantum properties of
the resulting state would still remain or diminish under such an
assumption.

Recently, such an 
analogy of Schr\"odinger's cat paradox, where
the state corresponding to the virtual cat
 is a significantly mixed thermal state, was investigated \cite{jr06}.
A thermal state with a high temperature is considered a classical state
in quantum optics.
As the temperature of the thermal state increases, the degree of mixedness,
which can be quantified by linear entropy, rapidly approaches the maximum value.
When the temperature approaches infinity, the thermal state
does not show any quantum properties.
As a comparison, coherent states  with large amplitudes are
known as the most classical pure states \cite{Schr2},
and their superposition 
is often regarded as a superposition of classical states
\cite{WScat}.
However, coherent states are still pure states which may not well represent truly
classical systems, and
they display some nonclassical features \cite{Johansen}.
In Ref.~\cite{jr06}, it was
shown that prominent quantum properties can actually be transferred
from a microscopic superposition to a significantly mixed thermal state
(i.e. a thermal state of which the degree of mixedness is close to the maximum value)
at a high temperature through an experimentally feasible process.
This result clarifies that 
unavoidable initial mixedness of the cat
does not preclude strong quantum phenomena.

One of the results in Ref.~\cite{jr06}
is that quantum entanglement can be produced 
between thermal states with nearly
the maximum Bell-inequality violation
when the temperatures of both modes goes to infinity.
In previous related results,
Bose {\it et al.} showed that entanglement can arise when two systems interact
if one of the system are pure even when the other system is extremely mixed \cite{Bose}.
There is an interesting previous
example shown by Filip {\it et al.} for the maximum violation of Bell's 
inequality when one of the modes
is an extremely mixed thermal state \cite{Filip}.
Very recently, Ferreira {\it et al.} showed that entanglement can be generated
at any finite temperature between high Q cavity mode
field and a movable mirror thermal state \cite{FV}.
However, in these example \cite{Bose,FV,Filip} 
only one of the modes is considered a large thermal state \cite{Bose,FV,Filip} 
and entanglement vanishes in the infinite temperature limit \cite{Bose,FV},
which is obviously in contrast to the result presented in Ref.~\cite{jr06}.
Entanglement for both of the modes at the thermal limit of the infinitely high temperature
has not been found before.
Remarkably, the violation of Bell's inequality in our examples
reaches up to Cirel'son's bound \cite{C80}
even in this infinite-temperature limit for both modes.
As Vedral \cite{Vedral} and Ferreira {\it et al.} \cite{FV} pointed out 
it is believed that high temperatures reduce entanglement and all entanglement
vanishes if the temperature is high enough, which is obviously not the case
in Ref.~\cite{jr06}.

The purpose of this paper is twofold.
Firstly, we review and further investigate 
various properties of superpositions and entanglement
of thermal states at high temperatures
\cite{jr06}.
In particular, we investigate two classes of 
highly mixed symmetric states in the phase space.
Both the classes of these states do not show typical interference
patterns in the phase space while they manifest
strong singular behaviors.
Interestingly, the first class of states
has neither squeezing properties 
nor negative values in their Wigner functions,
however, they are found to be highly nonclassical states.
The second class of states has the maximum negativity
in the Wigner function.
Further, we discuss the possibility of quantum information processing
with thermal-state qubits.
We introduce thermal-state qubits and thermal-Bell states,
which are a generalization of pure Bell states.
We show that four thermal-Bell states
can be well discriminated by nonlinear interactions
without photon number resolving measurements.
Quantum teleportation and gate operations
for thermal-state qubits can be realized
using the Bell measurement scheme.

This paper is organized as follows. 
In Sec.~II, we review the generation process
of superpositions of thermal states
and study their characteristics. 
In Sec.~III, we study entanglement of thermal states,
i.e., Bell inequality violations.
In Sec.~IV, we discuss the possibility of
quantum information processing using thermal states.
We first define the thermal-state qubit and the Bell-basis states
using thermal-state entanglement. We then show that the four
Bell states can be well discriminated by homodyne detection and
two Kerr nonlinearities. It follows that quantum teleportation
and quantum gate operations can be realized with thermal-state
qubits. We conclude with final remarks in Sec.~V.

\section{Superpositions of thermal states}

\subsection{Generation of thermal-state superpositions}
 
Let us first consider a two-mode harmonic oscillator system.
A displaced thermal state can be defined as
\begin{equation}
\rho^{th}(V,d)=\int d^2\alpha P^{th}(V,d)
|\alpha\rangle\langle\alpha|
\end{equation}
where $|\alpha\rangle$ is a coherent state of amplitude $\alpha$ and
\begin{equation}
P_\alpha^{th}(V,d)=\frac{2}{\pi(V-1)}
\exp[-\frac{2|\alpha-d|^2}{V-1}]
\end{equation}
with variance  $V$ and displacement $d$ in the phase space.
The thermal temperature $\tau$ increases as $V$ increases
as $e^{\hbar\nu/\tau}=(V+1)/(V-1)$, where $\hbar$ is Planck's constant and
$\nu$ is the frequency \cite{Walls}.
Suppose that a microscopic superposition state
\begin{equation}
|\psi\rangle_a=\frac{1}{\sqrt{2}}(|0\rangle_a+|1\rangle_a),
\label{ms}
\end{equation} 
where $|0\rangle$ and $|1\rangle$ are the ground and first excited states
of the harmonic oscillator,
 interacts with a thermal state $\rho^{th}_b(V,d)$ and 
the interaction Hamiltonian is
\begin{equation}
{\cal H}_K= \lambda \hat a^\dagger \hat a \hat b^\dagger \hat b
\label{CK}
\end{equation}
which corresponds to the cross Kerr nonlinear interaction. 
The resulting state is then
\begin{equation}
\label{tmc}
\begin{aligned}
\rho^{ent}_{ab}=&\frac{1}{2}
\int d^2\alpha P^{th}(V,d)
\Big\{
|0\rangle\langle 0|\otimes|\alpha\rangle\langle\alpha| \\
&+|1\rangle\langle 0|\otimes|\alpha e^{i\varphi}\rangle\langle\alpha|
+|0\rangle\langle 1|\otimes|\alpha\rangle\langle\alpha e^{i\varphi}|\\
&+|1\rangle\langle 1|\otimes|\alpha
e^{i\varphi}\rangle\langle\alpha e^{i\varphi}|
\Big\}
\end{aligned}
\end{equation}
and $\varphi$ is determined by the strength of the nonlinearity
$\lambda$ and the interaction time.
The Wigner representation of $\rho^{ent}_{ab}$ is
\begin{widetext}
\begin{equation}
W^{ent}_{ab}(\alpha,\beta)=\frac{1}{\pi}e^{-2|\alpha|^2}\Big\{
W^{th}(\beta;d)+2\alpha V^{c}(\beta;d)
+2[\alpha V^{c}(\beta;d)]^*+(4|\alpha|^2-1)W^{th}
(\beta;de^{i\varphi})\Big\}
\label{went}
\end{equation}
where $\alpha$ and $\beta$ are complex numbers parametrizing the phase spaces
of the microscopic and macroscopic systems respectively and
\begin{eqnarray}
&&W^{th}(\alpha;d)=\frac{2}{\pi V}
\exp[-\frac{2|\alpha-d|^2}{V}],\\
&&V^{c}(\alpha;d)=\frac{2}{\pi J K}
\exp[-\frac{2}{K}(1-e^{i \varphi})d^2  
-\frac{1}{J}
(\alpha-\frac{2e^{i\varphi}d}{K})(\alpha^*-\frac{2d}{K})],
\label{eq:mix-density}
\end{eqnarray}
$K=2+(V-1)(1-e^{i\varphi})$,
$J=(\sin\varphi/2+iV\cos\varphi/2)/(2V\sin\varphi/2+2i\cos\varphi/2)$,
and $d$ has been assumed real without loss of generality.
If one traces $\rho_{ab}^{ent}$ over mode $a$, the remaining state
will be simply in a classical mixture of two thermal states
and its Wigner function will be positive everywhere.
However, if one measures out the ``microscopic part'' 
on the superposed basis, i.e.,
$(|0\rangle_a\pm|1\rangle_a)/\sqrt{2}$,
the ``macroscopic part'' for mode $b$ 
may not lose its nonclassical characteristics.
Such a measurement on the the superposed basis 
will reduce the remaining state  to
\begin{equation}
\label{smc0}
\rho^{sup(\pm)}=
{\cal N}_s^\pm
\int d^2\alpha P^{th}(V,d)
\Big\{
|\alpha\rangle\langle\alpha|
\pm|\alpha e^{i\varphi}\rangle\langle\alpha|
\pm|\alpha\rangle\langle\alpha e^{i\varphi}|
+|\alpha e^{i\varphi}\rangle\langle\alpha e^{i\varphi}|
\Big\},
\end{equation}
where ${\cal N}_s^\pm$ are the normalization factors, and
its Wigner function is
\begin{equation}
W^{sup(\pm)}(\alpha)
={\cal N}_s^\pm\{W^{th}(\alpha;d)\pm V^{c}(\alpha;d)
\pm\{V^{c}(\alpha;d)\}^*+W^{th}(\alpha;de^{i\varphi})\}.
\label{swf}
\end{equation}
\end{widetext}
The $\pm$ signs in Eqs.~(\ref{eq:mix-density})
and (\ref{smc0}) correspond to the two possible results from
the measurement of the microscopic system.
The state in Eq.~(\ref{swf}) is a superposition of   
two thermal states.

A feasible experimental setup
to generate superpositions of thermal states
is atom-field interactions in cavities,
where a $\pi/2$ pulse can be used to prepare the atom in a superposed state.
This type of experiment has already been performed to produce
a superposition of coherent states \cite{Tu}.
In our cases, simply thermal states can be used instead of
coherent states.
 Another possible setup is
an all-optical scheme with free-traveling fields and a cross-Kerr medium,
where a standard single-photon qubit
could be used as the microscopic superposition.
Recently, there have been theoretical and experimental efforts to
produce and observe giant Kerr nonlinearities 
using electromagnetically induced transparency \cite{Hau}.
Furthermore, it was shown that a weak Kerr nonlinearity can still be
useful if a initially strong field is employed in this type of experiment
\cite{Jeong05}. We shall further explain this with examples in Sec.~III.

\subsection{Negativity of the Wigner function}

The negativity of the Wigner function is known as an indicator of
non-classicality of quantum states.
In order to observe negativity of the Wigner function
in a real experiment, its absolute minimum negativity
should be large enough.
The minimum negativity of the Wigner function in Eq.~(\ref{went})
for $V=1$
 is $-0.144$ for $d=0$ and $-0.246$ for $d\rightarrow\infty$.
Now suppose the initial state can be considered a classical thermal state
by letting $V\gg1$. 
One might expect that the negativity would be washed out as the initial state becomes mixed, but 
this is not the case. The minimum negativity
actually increases as $V$ gets larger. 
If $V\rightarrow\infty$, 
the minimum negativity of the Wigner function
(\ref{went}) is $-0.246$
regardless of $d$:
no matter how mixed the initial thermal state was, 
the minimum 
negativity of Wigner function is found to be a large value.
The point in the phase space which gives the minimum negativity
when $V\gg1$ or $d\gg0$ is $(-\frac{1}{2},0)$ and has negativity
\begin{equation}
W_{neg}\equiv W_{ab}^{ent}(-\frac{1}{2},0)
=\frac{2(-2+\frac{1}{V}\exp[-\frac{2d^2}{V}])}{\pi^2\sqrt{e}}.
\end{equation}
It can be shown that $W_{neg}$ approaches 
$-4/(\pi^2\sqrt{e})\approx-0.246$
when either $d\rightarrow\infty$ or $V\rightarrow\infty$.

This effect is obviously due to the interaction between the microscopic
superposition and the macroscopic thermal state.
If the initial microscopic state is not superposed, e.g., 
$|\psi\rangle_a=|1\rangle_a$,
the resulting state will be a simple direct product,
$(|1\rangle\langle 1|)_a\otimes\rho^{th}_b(V,-d)$.
Whilst for $V=1$ this state will exhibit negativity, this is washed out and tends to zero as 
$V\rightarrow\infty$.
Needless to say, if it was $|0\rangle_a$ instead of $|1\rangle_a$,
the resulting Wigner function will be 
a direct product of two Gaussian states 
whose Wigner fucntion can never be negative.
The superpositon state 
(\ref{ms}) 
plays the crucial role in making
the minimum negativity of the resulting Wigner function
always saturate to a certain negative value 
 no matter how mixed and
classical the initial state of the other mode becomes.

The Wigner functions of the single-mode states, $W^{sup(\pm)}(\alpha)$,
in Eq.~(\ref{swf}) show large negative values. 
The minimum negativity of the Wigner function $W^{sup(-)}(\alpha)$
is $W^{sup(-)}(0)=2/\pi$ regardless of the values of $V$ and $d$.
On the other hand, the minimum negativity of the Wigner function
$W^{sup(+)}(\alpha)$ approaches $2/\pi$ for $d\rightarrow\infty$
and disappears when $d=0$.

\begin{figure}
\centerline{\scalebox{0.48}{\includegraphics{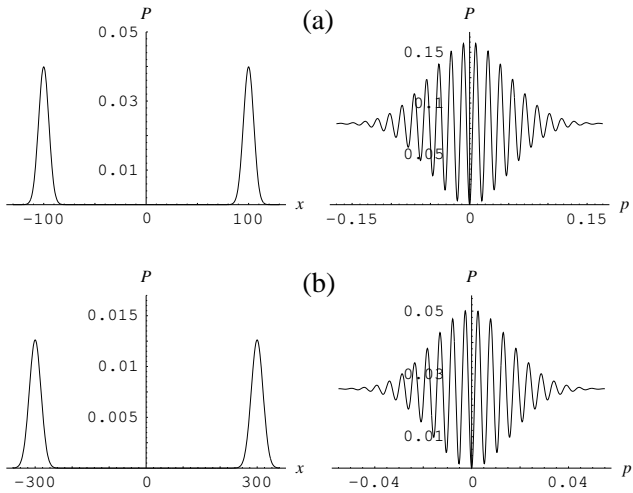}}}
\caption{The probability distributions of $x$ (left) and $p$
(right) for a ``superposition'' of 
two distant thermal states. A thermal state with a large mixedness 
is converted to such a ``thermal-state superposition'' 
by interacting with a microscopic superpotion (see text).
The variance $V$ and displacement $d$
for the thermal state are chosen as (a) $V=100$ and $d=100$,
and (b) $V=1000$ and $d=300$.
The fringe visibility is 1 regardless of $V$ and
the fringe spacing (the distance between the fringes)
does not depend on the variance (i.e. mixedness) but
only on the distance $d$ between the two component thermal states. 
}
\label{fig1}
\end{figure}

\subsection{Quantum interference in the phase space}

When $\varphi=\pi$,
the state (\ref{smc0}) becomes
\begin{equation}
\rho^\pm=N(\rho^{th}(V,d)\pm\sigma(V,d)\pm\sigma(V,-d)+\rho^{th}(V,-d)),
\end{equation}
where 
$\sigma(V,d)=\int d^2\alpha P^{th}(V,d)|-\alpha\rangle\langle\alpha|$
and
\begin{equation}
N=2\big(1\pm\frac{\exp[-\frac{2d^2}{V}]}{V^2}\big).
\label{e:N}
\end{equation}
If the initial state for mode $b$ is a pure coherent state,
i.e., $V=1$, the measurement on the superposed basis 
for mode $a$ will produce 
a superposition of two pure coherent states as
\begin{equation}
|\widetilde\Psi_\pm\rangle
=\frac{1}{\sqrt{1\pm e^{-2|\alpha|^2}}}(|\alpha\rangle\pm|-\alpha\rangle),
\end{equation}
where $\alpha=d$.
The probability ${\cal P}_\pm$ to obtain the state $\rho^\pm$
 is obtained as \cite{ref19}
\begin{equation}
{\cal P}_\pm=
\langle\psi^\pm|{\rm Tr}_b[\rho^{ent}_{ab}]|\psi^\pm\rangle
=\frac{1}{2}(1\pm\frac{\exp[-\frac{2 d^2}{V}]}{V}),
\label{Pro}
\end{equation}
where $|\psi^\pm\rangle=(|0\rangle\pm|1\rangle)/\sqrt{2}$.
The probability 
approaches ${\cal P}_\pm=1/2$ when either $d$ or $V$
becomes large.

\begin{figure}
\centerline{\scalebox{0.45}{\includegraphics{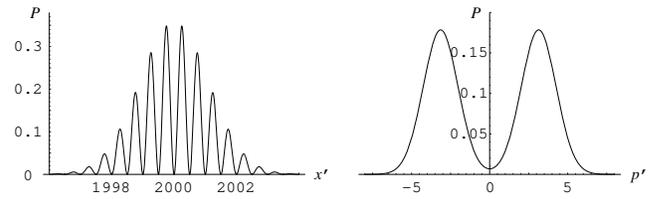}}}
\caption{The probability distributions $P$ 
for a ``superposition'' of thermal states where  $V=5$, $d=2000$,
$\varphi=\pi/1000$. 
The $x^\prime$ ($p^\prime$) axis 
in this figure has been rotated  by $\pi/2000$ from
the $x$ ($p$) axis for clarity.}
\label{fig4}
\end{figure}

As an analogy of Schr\"odinger's cat paradox,
the variance $V$ corresponds to the size the initial ``cat'', and
the distance $d$ between the two thermal component states corresponds 
to distinguishability between the ``alive cat'' and the ``dead cat''. 
Suppose that both $V$ and $d$ are very large for the initial thermal state.
The two thermal states $\rho^{th}(V,\pm d)$ become macroscopically
distinguishable when $d\gg\sqrt{V}$, and
our example may become
a more realistic analogy of the cat paradox in this limit.
Both the states $\rho^\pm$ in this case show
probability distributions with
two Gaussian peaks and interference fringes \cite{jr06}.  
Figure~\ref{fig1}
presents the probability distributions
of $x~(\equiv Re[\alpha])$ and $p~(\equiv Im[\alpha])$
for $\rho^-$  (a) when $V=100$ and $d=100$
and  (b) when $V=1000$ and $d=300$.
The probability distribution of $x$ ($p$) 
for $\rho^\pm$ can be obtained by
integrating the Wigner function of $\rho^\pm$ over $p$ ($x$).    
The two Gaussian peaks along the $x$ axis and
interference fringes along the $p$ axis shown in Fig.~\ref{fig1}
are a typical signature of a quantum superposition between 
macroscopically distinguishable states.
The visibility $v$ of the interference fringes is defined as
\cite{Walls}
\begin{equation}
v=\frac{I_{\max}-I_{\min}}{I_{\max}+I_{\min}},
\end{equation}
where $I=\int dx W^{sup(-)}(\alpha)$ and the maximum should
be taken over $p$.
It can be simply shown that the visibility $v$ is always 1 
regardless of the value of $V$. 
Note that $d$ should increase proportionally to $\sqrt{V}$ 
to maintain the condition of classical distingushability between
the two component thermal states $\rho^{th}(V,\pm d)$.
The interference fringes with high visibility
are incompatible with classical physics
and evidence of quantum coherence.
The fringe spacing (the distance between the fringes)
does not depend on $V$ but only on $d$, i.e.,
a pure superposition of coherent states shows
the same fringe spacing for a given $d$. 
We emphasize that 
the states shown in Fig.~\ref{fig1} are ``superpositions''
of severely mixed thermal states.

An experimental realization of a nonlinear effect
corresponding to $\varphi=\pi$ is very demanding
particularly in the presence of decoherence.
Here we point out that 
the method using a weak nonlinear effect
($\varphi\ll \pi$)
combined with a strong field 
($d\gg 1$) \cite{Jeong05} can
be useful to generate a thermal-state superposition
with prominent interference patterns.
In Fig.~\ref{fig4}, we have used experimentally accessible values, $V=5$, $d=2000$ and 
$\varphi=\pi/1000$, but the fringe visibility is still 1.
In this case, {\it decoherence during the nonlinear interaction
would be significantly reduced} because of 
the decrease of the interaction time
\cite{Jeong05}.
Note also that, if required, the state in Fig.~\ref{fig4} 
can be moved to the center of the phase space,
for example, using a biased  beam splitter (BS)
and a strong coherent field
\cite{Jeong05}.

\subsection{Symmetric macroscopic quantum states}

Let us assume that $d=0$, i.e., the initial state
is the thermal state, $\rho^{th}(V,0)$,
at the origin of the phase space. 
In this case, the thermal-state superpositions, $\rho^\pm$,
are 
produced with probabilities, ${\cal P}_\pm=(1/2)\{1\pm(1/V)\}$,
respectively.
Figure~\ref{fig:sym} shows the Wigner functions of  $\rho^+$
dependent on the interaction time between the macroscopic thermal state
and the microscopic superposition in a cross Kerr medium.
The state is always symmetric in the phase space
regardless of the interaction time as shown in Fig.~\ref{fig:sym}.
In this figure, the initial state is a thermal state of $V=100$ (Fig.~\ref{fig:sym}(a)).
In a relatively short time ($\theta=\pi/32$ and $\theta=\pi/16$),
the state shows some interference patterns.
When $\theta=\pi$, the evolved
state looks very localized around the origin
as shown in Fig~\ref{fig:sym}.
The generated state at $\theta=\pi$ does not show negativity
of the Wigner function nor squeezing properties.
On the other hand, a well defined $P$ function does not exist
for this state.

In the case of $\rho^-$, with the same assumption $d=0$,
the Wigner function at
$\varphi=\pi$
has the minimum negativity ($-2/\pi$) at the origin
regardless of $V$.
As a result of the interaction with the microscopic
superposition, a deep hole 
to the negative direction below zero
has been formed around the origin for $\rho^-$
as shown in Fig.~\ref{fig:sym2} .

\begin{widetext}
\begin{center}
\begin{figure}
\centerline
{~~~~~~(a)~~~~~~~~~~~~~~~~~~~~~~~~~~~~~~~~~~~~~~~~~~~~~~~
(b)~~~~~~~~~~~~~~~~~~~~~~~~~~~~~~~~~~~~~~~~~~~~~~~~~
(c)~~~~~~}
\centerline{\scalebox{0.56}{\includegraphics{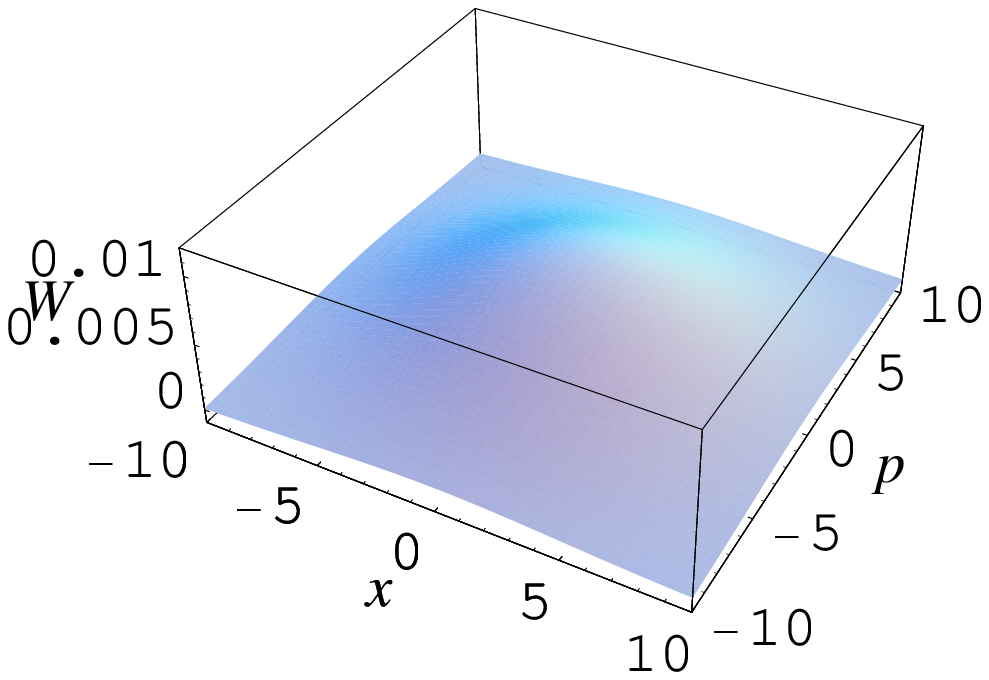}}
\scalebox{0.56}{\includegraphics{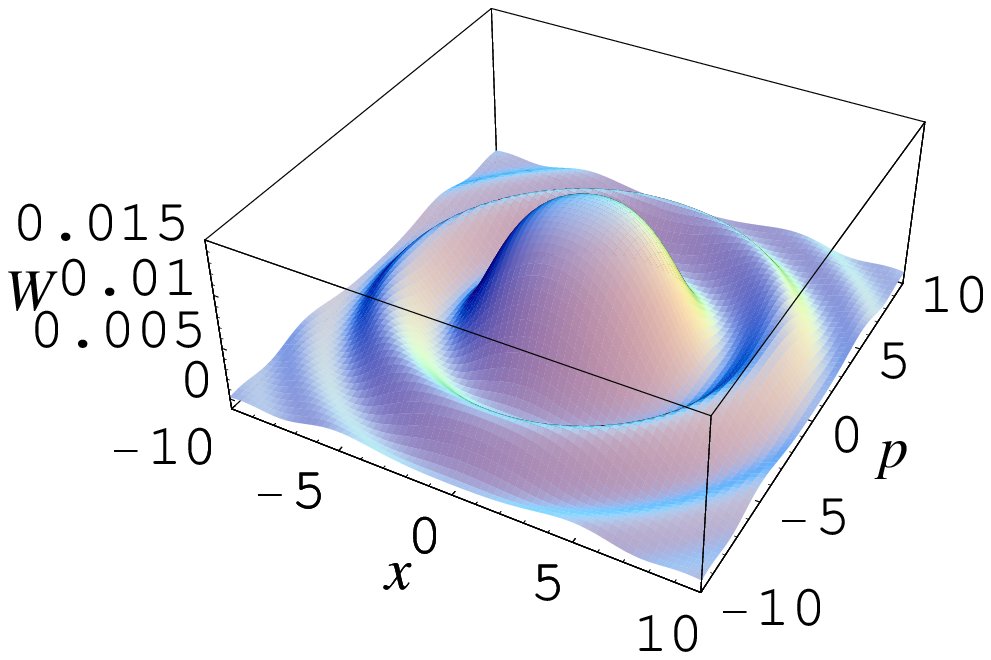}}
\scalebox{0.56}{\includegraphics{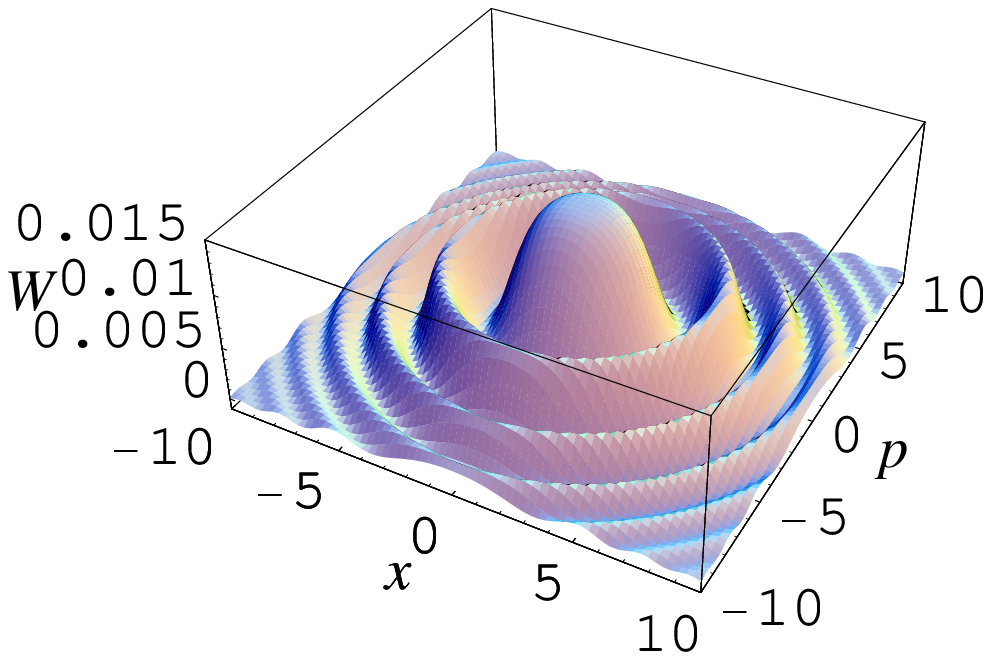}}}
\vspace{0.03cm}
\centerline
{~~~~~~(d)~~~~~~~~~~~~~~~~~~~~~~~~~~~~~~~~~~~~~~~~~~~~~~~
(e)~~~~~~~~~~~~~~~~~~~~~~~~~~~~~~~~~~~~~~~~~~~~~~~~~
(f)~~~~~~}
\centerline{\scalebox{0.56}{\includegraphics{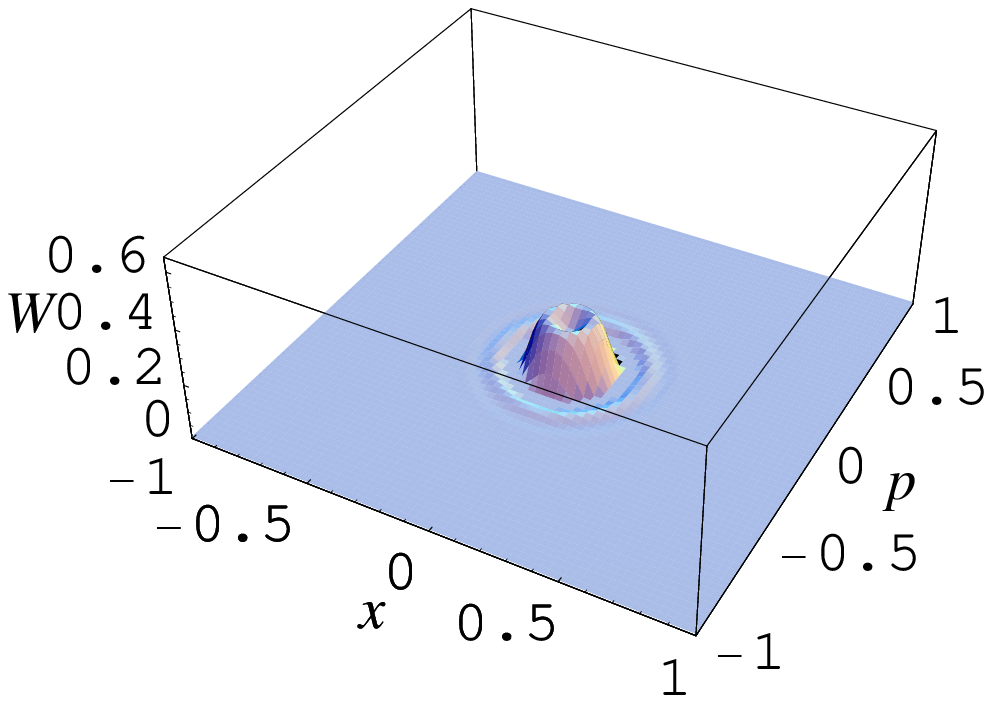}}
\scalebox{0.56}{\includegraphics{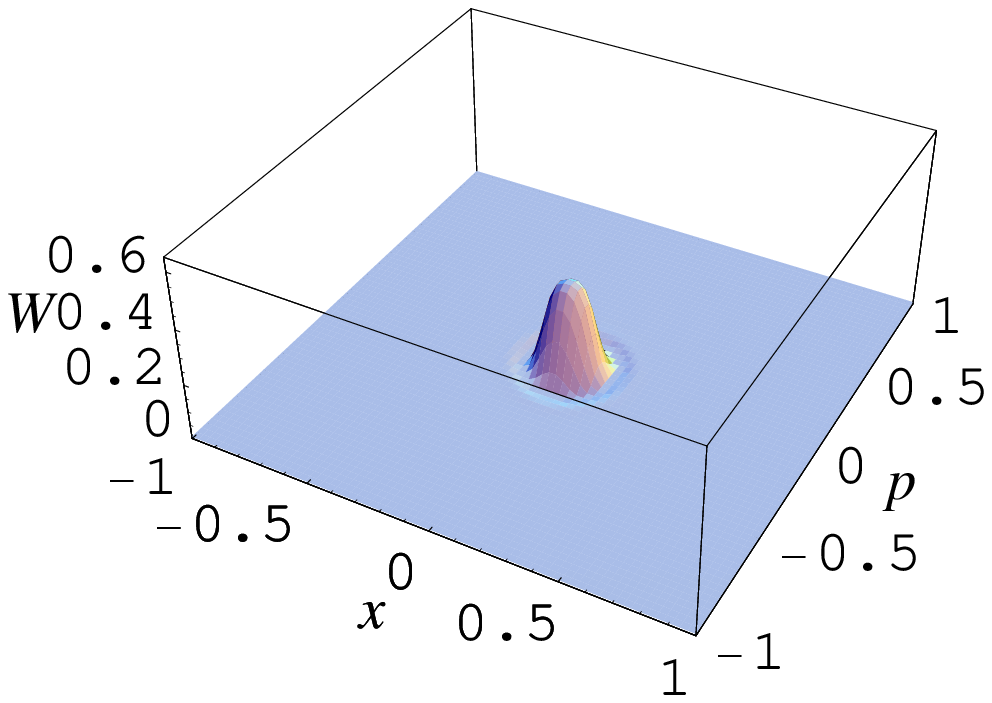}}
\scalebox{0.56}{\includegraphics{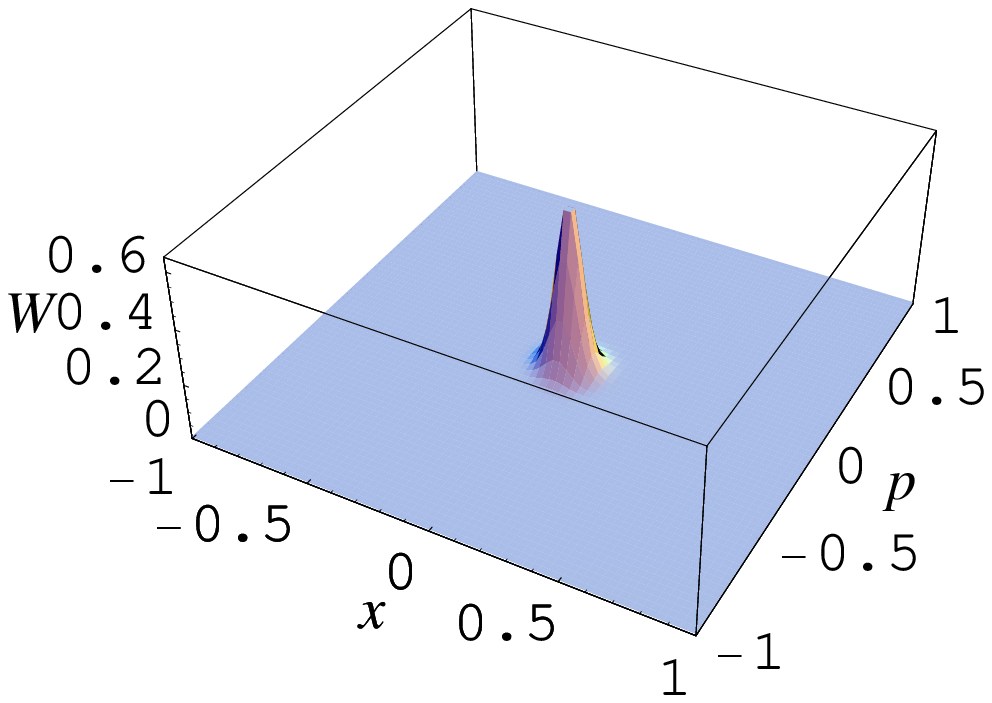}}}
\caption{(Color online) 
The time dependent Wigner functions of the thermal state of $V=100$
at the origin ($d=0$) after an interaction with
a microscopic superposition and a conditional measurement.
The measurement result on the microscopic part was supposed 
to be $(|0\rangle+|1\rangle)/\sqrt{2}$.
The interaction times are
(a) $\theta=\lambda t=0$, (b)  $\theta=\lambda t=\pi/32$, (c) $\theta=\pi/16$,
(d) $\theta\approx 3.102$,
(e) $\theta\approx 3.122$
and (f) $\theta=\pi$.}
\label{fig:sym}
\end{figure}
\end{center}
\end{widetext}

\begin{figure}
\centerline{\scalebox{0.6}{\includegraphics{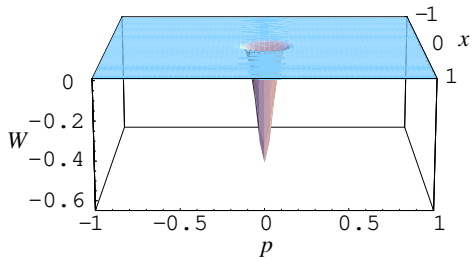}}}
\caption{(Color online) The Wigner function of the thermal state of $V=100$
at the origin ($d=0$) after an interaction with
a microscopic superposition and a conditional measurement.
The measurement result on the microscopic part was supposed 
to be $(|0\rangle-|1\rangle)/\sqrt{2}$
with the interaction time $\theta=\lambda t=\pi$.}
\label{fig:sym2}
\end{figure}

\section{Entanglement between thermal states}

Entanglement between macroscopic objects and 
its Bell-type inequality tests
are an important issue.
In this section, we shall show that entanglement can be generated
between high-temperature thermal states
even when the temperature of each mode goes to infinity.

\subsection{Entanglement using two initial thermal states}

\begin{figure}
\centerline{\scalebox{0.55}{\includegraphics{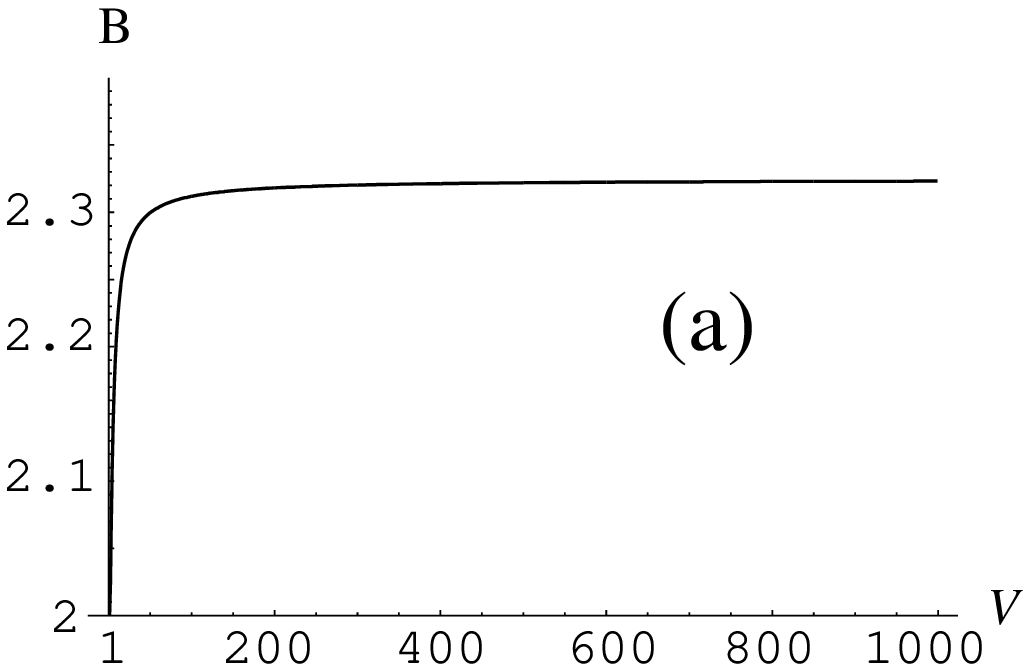}}}
\vspace{0.3cm}
\centerline{\scalebox{0.55}{\includegraphics{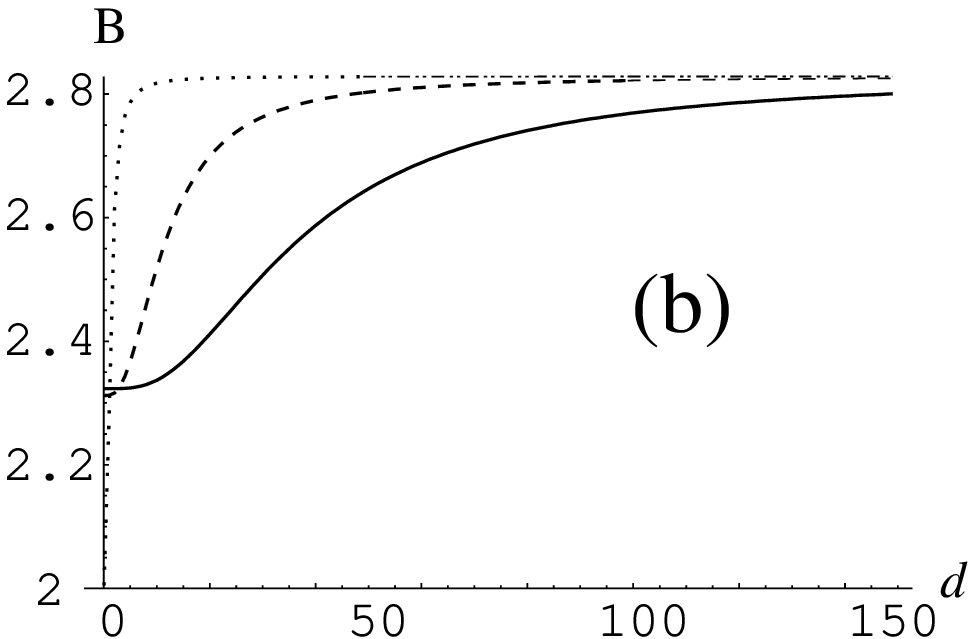}}}
\caption{(a) The optimized violation, ${\rm B}\equiv|B^+|_{max}$,
of Bell-CHSH inequality for the
``thermal-state entenglement'', $\rho_+$, of $V=1000$ (solid curve)
and $V=100$ (dashed curve). The Bell-violation of a pure entangled
coherent state, i.e., $V=1$, has been plotted for comparison (dotted curve).
The Bell-violation B approaches its maximum bound,
$2\sqrt{2}$, when $d\gg\sqrt{V}$ 
regardless of the level of the mixedness $V$.
(b) The optimized Bell-violation B against $d$ for the different
type of thermal-state entanglement
generated using a 50:50 beam splitter from $\rho^+$. 
$V=1000$ (solid curve), $V=100$ (dashed curve)
and $V=1$ (dotted curve).
}
\label{bell}
\end{figure}

If the microscopic superposition interacts with two thermal states,
$\rho_b^{th}(V,d)$ and $\rho_c^{th}(V,d)$,
and the microscopic particle is measured out on the superposed basis,
 the resulting state will be
\begin{equation}
\begin{aligned}
&\rho^{tm(\pm)}
=N_t\big\{\rho^{th}(V,d)\otimes\rho^{th}(V,d)
\pm\sigma(V,d)\otimes\sigma(V,d)\\
&\pm\sigma(V,-d)\otimes
\sigma(V,-d)+\rho^{th}(V,-d)\otimes\rho^{th}(V,-d)\big\}
\end{aligned}
\label{e-tm}
\end{equation}
where 
\begin{equation}
N_t=2\big(1\pm\frac{\exp[-\frac{4d^2}{V}]}{V^2}\big).
\label{nmt}
\end{equation}
Such two-mode thermal-state entanglement can be generated using
two cavities and an atomic state detector \cite{mskim}.
Extending the two cavities to $N$ cavities, entanglement of
$N$-mode thermal states can also be generated. Such a state
is an analogy of the $N$-mode pure GHZ state \cite{GHZ}
but each mode is extremely mixed.
Here we shall consider the Bell-CHSH inequality \cite{Bell,CHSH} 
with photon number parity measurements \cite{mskim,BW}. 
The parity measurements can be performed in
a high-Q cavity using a far-off-resonant
interaction between a two-level atom and the field \cite{eh}.
The Bell-CHSH inequality can be
represented in terms of the Winger function as
\cite{BW}
\begin{equation}
\begin{aligned}
&|B^{(\pm)}|=\frac{\pi^2}{4}|W^{tm(\pm)}(\alpha,\beta)
+W^{tm(\pm)}(\alpha,\beta^\prime)\\
&~~~~~~~~~~+W^{tm(\pm)}(\alpha^\prime,\beta)
-W^{tm(\pm)}(\alpha^\prime,\beta^\prime)|\leq2,
\end{aligned}
\end{equation}
where $W^{tm(\pm)}(\alpha,\beta)$ is the Wigner function of
$\rho^{tm(\pm)}$ in Eq.~(\ref{e-tm}).
As shown in Fig.~\ref{bell}, the Bell-violation
approaches the maximum bound for a bipartite measurement,
$2\sqrt{2}$ \cite{C80}, when $d\gg\sqrt{V}$ 
regardless of the level of the mixedness $V$,
i.e., the temperatures of the thermal states.
Note that it is true for both of $\rho_+$ and $\rho_-$ even though only the case
of $\rho_+$ has been plotted in Fig.~\ref{bell}(a). 
This implies that entanglement of nearly 1 ebit has been produced 
between the two significantly mixed thermal states for
$d\gg\sqrt{V}$, and such ``thermal-state entanglement''
cannot be described by a local theory.

\subsection{Entanglement using a beam splitter}

\begin{figure}
\centerline{\scalebox{0.55}{\includegraphics{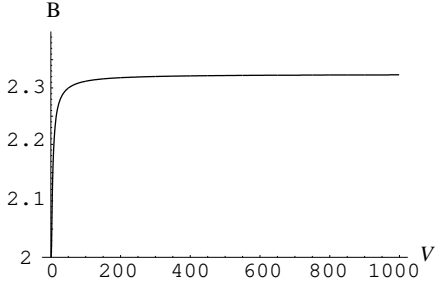}}}
\caption{
The optimized Bell-violation B against $V$ for the slightly different
type of thermal-state entanglement
generated using a 50:50 beam splitter 
using $\rho^+$ when $d=0$. 
}
\label{bell2}
\end{figure}

A different type of macroscopic 
entanglement can be generated
by applying the beam splitter operation
\begin{equation}
\exp[\theta/2(e^{i\phi}
\hat{a}^{\dagger}_{s}\hat{a}_{d}-e^{-i\phi}
\hat{a}_{d}^{\dagger}\hat{a}_{s})],
\end{equation}
on the ``thermal-state superpositions'' in Eq.~(\ref{smc0}).
The state after passing through a 50:50 beam splitter can be represented as
\begin{widetext}
\begin{equation}
N\int d^2\alpha P^{th}_\alpha(V,d)
\Big(|\frac{\alpha}{\sqrt{2}},-\frac{\alpha}{\sqrt{2}}\rangle
\pm|-\frac{\alpha}{\sqrt{2}},\frac{\alpha}{\sqrt{2}}\rangle\Big)
\Big(\langle\frac{\alpha}{\sqrt{2}},-\frac{\alpha}{\sqrt{2}}|
\pm\langle -\frac{\alpha}{\sqrt{2}},\frac{\alpha}{\sqrt{2}}|\Big),
\end{equation}
\end{widetext}
where $N$ is defined in Eq.~(\ref{e:N}).
When $d$ is large, this state
violates the Bell-CHSH inequality to the maximum bound $2\sqrt{2}$
regardless of the level of mixedness $V$ as shown in Fig.~\ref{bell}(b).
Again, it is true for both of $\rho_+$ and $\rho_-$ even though only the case
of $\rho_+$ has been plotted in Fig.~\ref{bell}(b). 
Furthermore, these states severely violate Bell's inequality even when $d=0$ as
$V$ increases as shown in Fig.~\ref{bell2}.
We have found that the optimized Bell violation of these states
approaches $2.32449$ for $V\rightarrow\infty$.
Interestingly, this value is exactly the same as the optimized  Bell-CHSH violation
for a pure two-mode squeezed state in the infinite squeezing limit \cite{jeongsonkim}.
Note that multilmode entangled states can be generated using multiple beam splitters.

\begin{figure}
\centerline{\scalebox{0.55}{\includegraphics{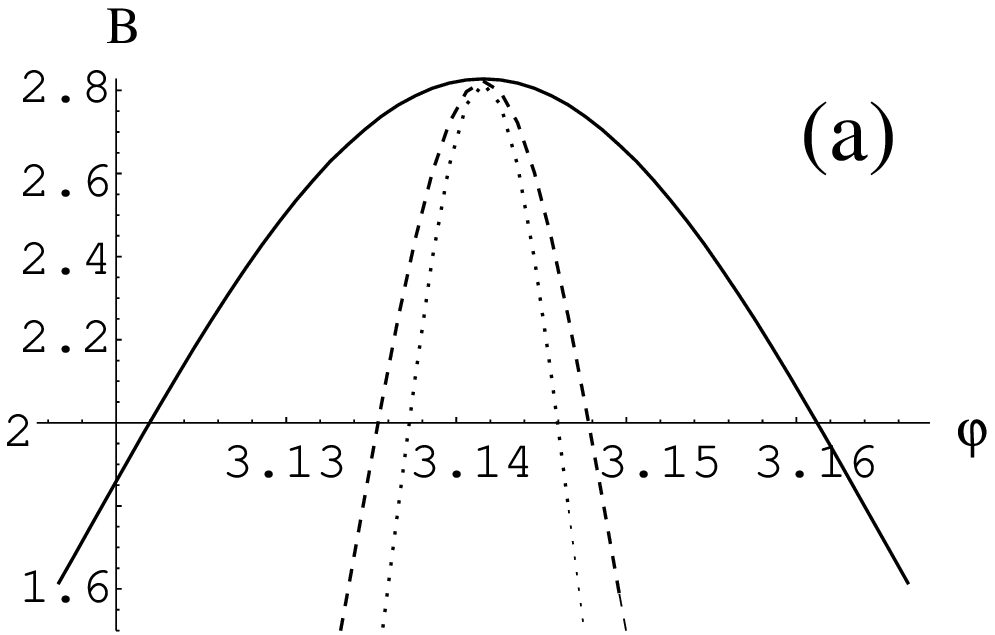}}}
\vspace{0.3cm}
\centerline{\scalebox{0.55}{\includegraphics{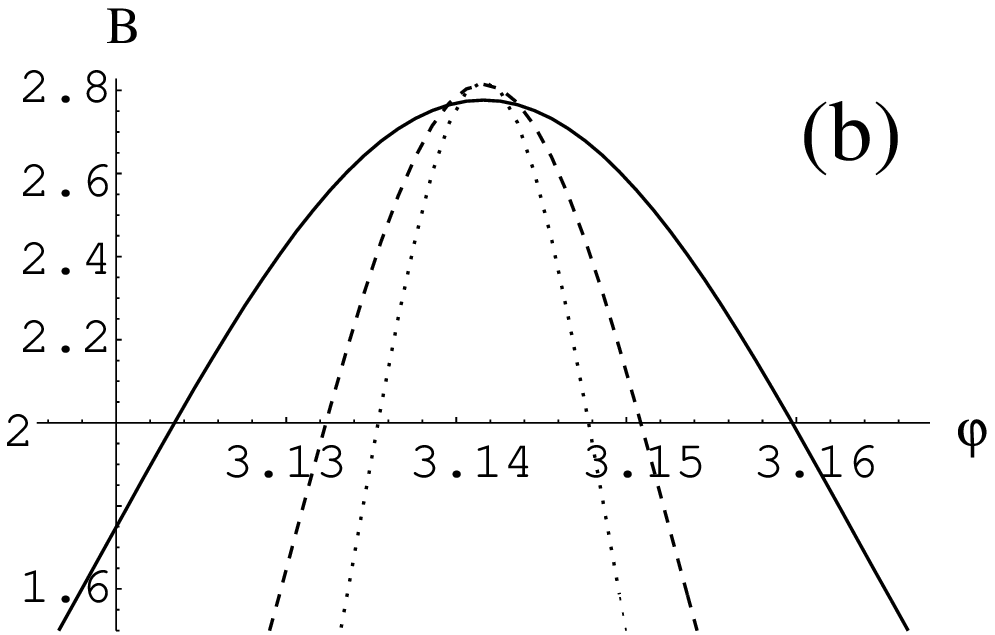}}}
\caption{(a) The Bell-CHSH function B
against $\theta$ ($=\lambda t$) for $V=1$ (solid curve),
$V=10$ (dashed curve) and $V=20$ (dotted curve)
for $d=30$.
(b)  The Bell-CHSH function for $d=10$ (solid curve),
$d=20$ (dashed curve) and $d=30$ (dotted curve)
for $V=10$.
The Bell violations
are more sensitive to the interaction time  
as either $V$ or $d$ increases.}
\label{fig:time}
\end{figure}

It should be noted that
 the Bell violations
are more sensitive to the interaction time  
when either $V$ or $d$ is larger.
Figure~\ref{fig:time} clearly shows this tendency.
Therefore, in order to observe the Bell violations
using the mixed state of $V$ (and $d$) large,
the interaction time in the Kerr medium
should be more accurate.

\section{Quantum information processing with thermal-state qubits}

In this section, we discuss the possibility of quantum
information processing with thermal-state qubits
and thermal-state entanglement.

\subsection{Qubits and Bell-state measurements}

We introduce a thermal-state qubit
\begin{equation}
\begin{aligned}
\rho^\psi
=|a|^2\rho^{th}(V,d)
\pm ab^*\sigma(V,d)
\pm a^*b\sigma(V,-d)+|b|^2\rho^{th}(V,-d),
\end{aligned}
\label{t-qubit}
\end{equation}
where $a$ and $b$ are arbitrary complex numbers.
The basis states, $\rho^{th}(V,d)$ and $\rho^{th}(V,-d)$,
can be well discriminated by a homodyne measurement
when $d$ is larger  than $V$.
The thermal state qubit (\ref{t-qubit})
can be re-written as
\begin{equation}
\rho^\psi=
\int d^2\alpha P^{th}_\alpha(V,d)
\big(a|\alpha\rangle+b|-\alpha\rangle\big)
\big(a^*\langle\alpha|+b^*\langle-\alpha|\big),
\label{t-qubit2}
\end{equation}
which can be understood as a generalization of
the coherent state qubit, 
$a|d\rangle+b|-d\rangle$,
where $|d\rangle$ is a coherent state of amplitude
$d$.
The thermal-state qubit (\ref{t-qubit2})
becomes identical to 
the coherent-state qubit when $V=1$.

We also define four thermal-Bell states as
\begin{widetext}
\begin{eqnarray}
&&\rho^{\Phi(\pm)}
=N_t\big\{\rho^{th}(V,d)\otimes\rho^{th}(V,d)
\pm\sigma(V,d)\otimes\sigma(V,d) 
\pm\sigma(V,-d)\otimes
\sigma(V,-d)+\rho^{th}(V,-d)\otimes\rho^{th}(V,-d)\big\}\\
&&\rho^{\Psi(\pm)}
=N_t\big\{\rho^{th}(V,d)\otimes\rho^{th}(V,-d)
\pm\sigma(V,d)\otimes\sigma(V,-d) 
\pm\sigma(V,-d)\otimes
\sigma(V,d)+\rho^{th}(V,-d)\otimes\rho^{th}(V,d)\big\}
\label{t-bell}
\end{eqnarray}
where $N_t$ was defined in Eq.~(\ref{nmt}).
The thermal-Bell states can be written as 
\begin{eqnarray}
\rho^{\Phi(\pm)}=
N_t\int d\alpha^2 d\beta^2 
P^{th}_\alpha(V,d)
P^{th}_\beta(V,d)
\big(|\alpha,\beta\rangle\pm|-\alpha,-\beta\rangle\big)
\big(\langle\alpha,\beta|\pm\langle-\alpha,-\beta|\big),\\
\rho^{\Psi(\pm)}=
N_t\int d\alpha^2 d\beta^2 
P^{th}_\alpha(V,d)
P^{th}_\beta(V,d)
\big(|\alpha,-\beta\rangle\pm|-\alpha,\beta\rangle\big)
\big(\langle\alpha,-\beta|\pm\langle-\alpha,\beta|\big).
\end{eqnarray}
\end{widetext}
For quantum information processing applications,
it is an important task to discriminate between the four Bell states.
Here we discuss two possible ways to discriminate 
between the thermal-Bell states 
(\ref{t-bell}). We shall only briefly describe the first scheme using
photon number resolving measurements 
and focus on the second scheme using nonlinear interactions.

\begin{figure}
\centerline{\scalebox{0.55}{\includegraphics{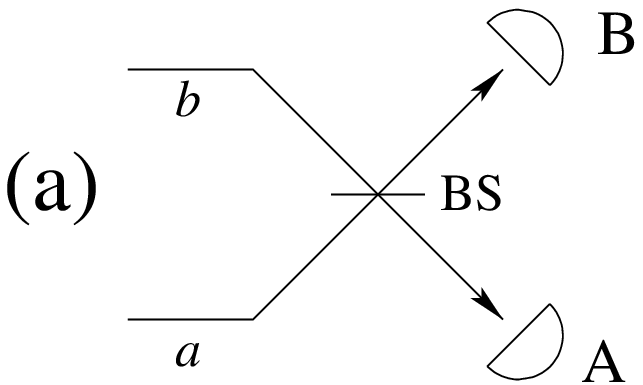}}}
\vspace{0.8cm}
\centerline{\scalebox{0.55}{\includegraphics{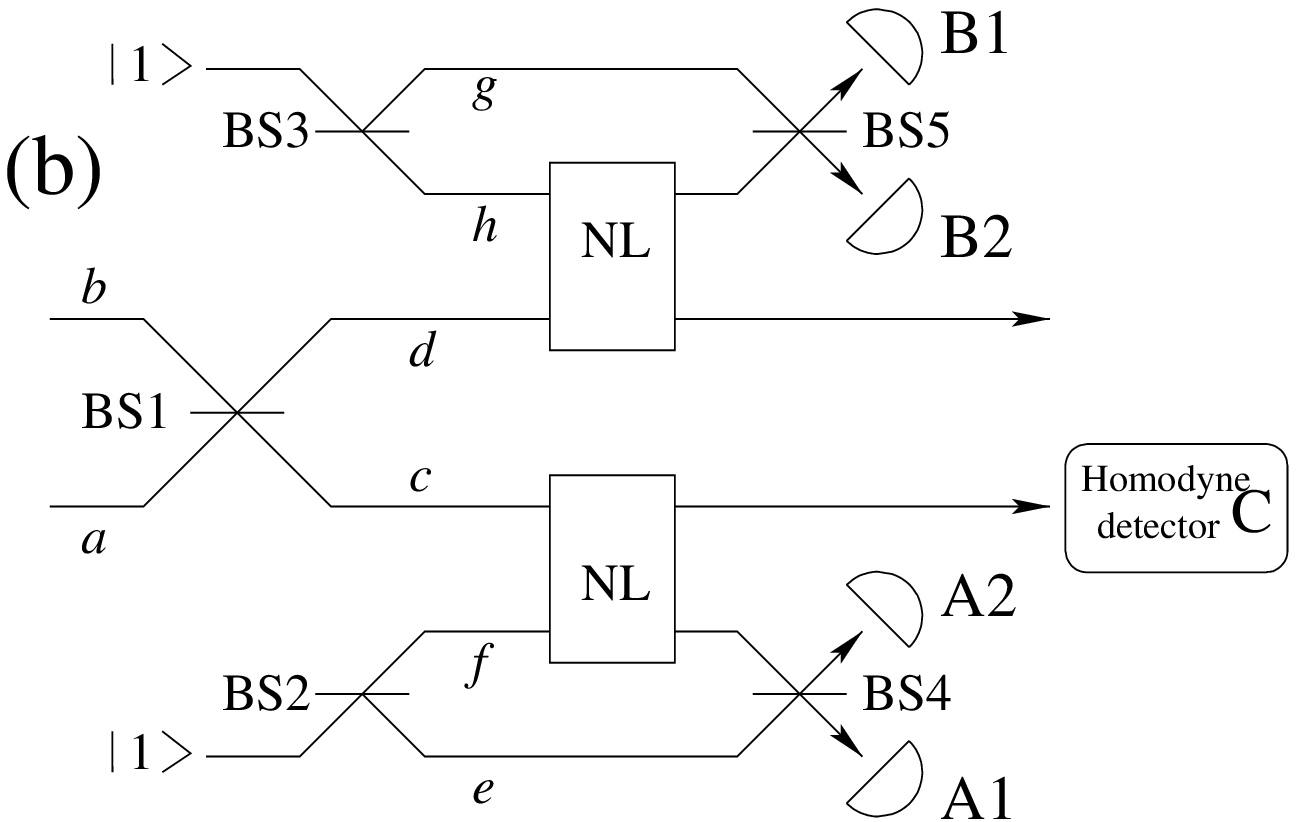}}}
\caption{A schematic of the thermal-Bell state measurement
(a) using  photon number resolving detection and (b) using
homodyne measurements with cross-Kerr nonlinear interactions (NL).
See text for details.}
\label{fig:scheme}
\end{figure}

\begin{figure}
\centerline{\scalebox{0.55}{\includegraphics{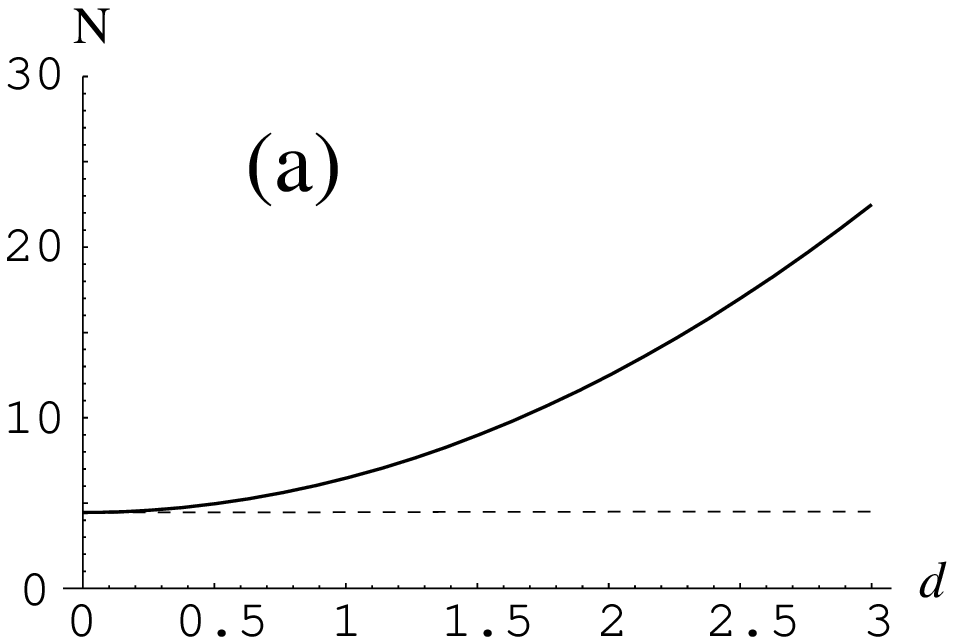}}}
\vspace{0.3cm}
\centerline{\scalebox{0.55}{\includegraphics{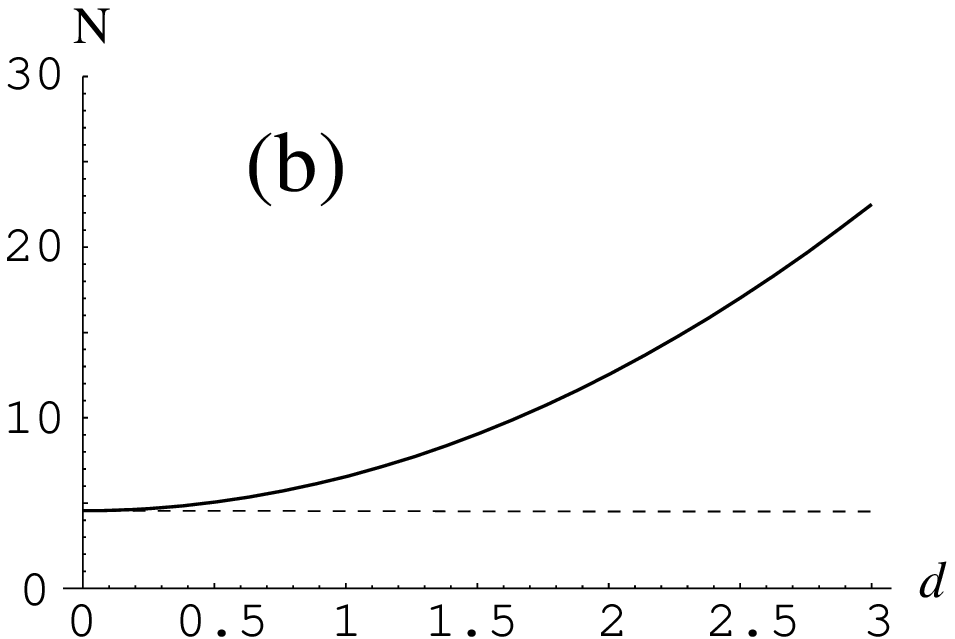}}}
\caption{The average photon number N
for the ``many-photon case'' (solid line)
and the ``few-photon case'' (dashed line) for $V=10$ against $d$
(a) when the input state is either $\rho^{\Phi(+)}$ 
or  $\rho^{\Psi(+)}$ 
and (b)  when the input state is either $\rho^{\Phi(-)}$ 
or  $\rho^{\Psi(-)}$.}
\label{fig:few-many}
\end{figure}

The first method is to simply use a 50-50 beam splitter
and two photon number resolving detectors as shown in Fig.~\ref{fig:scheme}(a).
This scheme is basically the same as the Bell-state measurement
scheme with pure entangled coherent states \cite{JKL01,Enk01}.
Let us suppose that the amplitude, $d$, is large enough, i.e.,
$d\gg \sqrt{V}$.
If the incident state was  $\rho^{\Phi(+)}$ or  $\rho^{\Phi(-)}$,
most of the photons are detected on detector A in in Fig.~\ref{fig:scheme}(a).
Meanwhile, most of the photons are detected on detector B 
when the incident state was $\rho^{\Psi(+)}$ or $\rho^{\Psi(-)}$.
The average photon numbers between the ``many-photon case''
and the ``few-photon case'' 
are compared in Fig.~\ref{fig:few-many}.
Furthermore, the states $\rho^{\Psi(+)}$ and $\rho^{\Phi(+)}$ contain
only even numbers of photons
while $\rho^{\Psi(-)}$ and $\rho^{\Phi(-)}$ contain only odd numbers of photons.
Therefore, all the four Bell states can be well discriminated by
analyzing numbers of photons detected at detectors A and B.
For example, if detector A detects many photons while detector B detects
few and the total photon number detected by the two detectors are even,
this means that state $\rho^{\Phi(+)}$ was measured by the thermal-Bell measurement.
The nonzero failure probability can be made arbitrarily small by
increasing $d$. 

However, the average photon numbers of the thermal-Bell states 
are high when $V\gg1$ and $d\gg1$.
In this case, it would be unrealistic 
to use photon number resolving detectors.
It would be an interesting question whether these four thermal-Bell states can be
distinguished by classical measurements, such as homodyne detection,
instead of photon number resolving detection.
Our alternative scheme employs cross-Kerr nonlinearities  
and single photon detectors as shown in Fig.~\ref{fig:scheme}(b).
Let us first suppose that the input field was $\rho^{\Phi(+)}$.
The incident two-mode state passes through a 50-50 beam splitter, BS1.
The state after passing through the 50:50 beam splitter, BS1, is
\begin{equation}
\begin{aligned}
\rho^{B}&=N_t\int d^2\alpha d^2\beta 
P^{th}_\alpha(V,d) P^{th}_\beta(V,d) \Big\{
|\eta,-\xi \rangle\langle \eta,-\xi |\\
&+|\eta,-\xi \rangle\langle -\eta,\xi | 
+|-\eta,\xi \rangle\langle \eta,-\xi |+
|-\eta,\xi \rangle\langle -\eta,\xi |\Big\}
\end{aligned}
\end{equation}
where $\eta=(\alpha+\beta)/\sqrt{2}$ and $\xi=(\alpha-\beta)/\sqrt{2}$.
Two dual-rail single photon qubits, $|\psi_+\rangle_{ee^\prime}$ and
$|\psi_+\rangle_{ff^\prime}$, where 
\begin{equation}
|\psi_+\rangle=\frac{1}{\sqrt{2}}(|0\rangle|1\rangle+|1\rangle|0\rangle),
\end{equation}
are prepared using two single photons and 50:50 beam splitters,
BS2 and BS3, as shown in
Fig.~\ref{fig:scheme}(b).
Then, traveling fields at modes $c$ and $d$
interacts with those of modes $e$ and $f$, respectively,
in cross-Kerr nonlinear media.
We suppose that the interaction time is $t=\pi/\lambda$, and
the resulting state is then
\begin{equation}
\rho^{B^\prime}=U_{ce} U_{df}\rho^{\cal B}_{c d} 
\rho^{q}_{ee^\prime}\rho^{q}_{ff^\prime}U_{ce}^\dagger U_{df}^\dagger
\label{imp}
\end{equation}
where $U_{ce}=\exp[i \pi {\cal H}^K_{ce} /\lambda\hbar]$
and $\rho^{q}=|\psi_q\rangle \langle\psi_q|$.
An explicit form of Eq.~(\ref{imp}) can then be simply obtained using
the identity
\begin{equation}
\begin{aligned}
&U_{ce}|\alpha\rangle_c|0\rangle_e=|\alpha\rangle_c|0\rangle_e,\\
&U_{ce}|\alpha\rangle_c|1\rangle_e=|-\alpha\rangle_c|1\rangle_e
\end{aligned}
\end{equation}
where $|\alpha\rangle$ is a coherent state.
However, we omit such an explicit expression in this paper for it is too lengthy.

After the nonlinear interactions,
the qubit parts, modes $e$, $e^\prime$, $f$ and $f^\prime$, should be measured with
the measurement basis 
\begin{equation}
\{|++\rangle,~
|+-\rangle,~
|-+\rangle,~
|--\rangle\}
\end{equation}
where 
$|++\rangle=|\psi_+\rangle_{ee^\prime}|\psi_+\rangle_{ff^\prime}$,
$|+-\rangle=|\psi_+\rangle_{ee^\prime}|\psi_-\rangle_{ff^\prime}$,
$|-+\rangle=|\psi_-\rangle_{ee^\prime}|\psi_+\rangle_{ff^\prime}$,
$|--\rangle=|\psi_-\rangle_{ee^\prime}|\psi_-\rangle_{ff^\prime}$,
and
$|\psi_-\rangle=(|0\rangle|1\rangle-|1\rangle|0\rangle)/\sqrt{2}$.
This measurement can be performed using
two 50:50 beam splitters, BS4 and BS5, and
four detectors, A1, A2, B1 and B2, as shown in Fig.~\ref{fig:scheme}(b).
If detector A1 and B1 click, i.e.,
the measurement result is 
$|++\rangle$,
the resulting state at modes $c$ and $d$ is
\begin{equation}
\begin{aligned}
\rho^{++}=&\frac{N_t}{4}\int d^2\alpha d^2\beta P^{th}_\alpha(V,d) P^{th}_\beta(V,d)\\
&~~~~~~~~\Big\{(|\eta\rangle+|-\eta\rangle)(\langle\eta|+\langle-\eta|)\Big\}_c\\
&~~~~\otimes\Big\{(|\xi\rangle+|-\xi\rangle)(\langle\xi|+\langle-\xi|)\Big\}_d
\label{ntp-1}
\end{aligned}
\end{equation}
Note that state $\rho^{++}$ 
is not normalized, which implies that the probability of obtaining the 
corresponding measurement result is not unity.  
The probability of obtaining this result is
\begin{equation}
P_{++}=\frac{(V+1)(V+e^{-\frac{4d^2}{V}})}{2(V^2+e^{-\frac{4d^2}{V}})}.
\end{equation}
When the result is 
either $|+-\rangle$ or $|-+\rangle$,
 the result is 
\begin{equation}
\langle\psi_2|\rho^{B^\prime}|\psi_2\rangle=
\langle\psi_3|\rho^{B^\prime}|\psi_3\rangle=0,
\end{equation}
which obviously means that the probability of the obtaining this
result is zero. When the result is $|--\rangle$, i.e.,
detector A2 and B2 click,
\begin{equation}
\begin{aligned}
\rho^{--}=&\frac{N_t}{4}\int d^2\alpha d^2\beta P^{th}_\alpha(V,d) P^{th}_\beta(V,d)\\
&~~~~~~~~\Big\{(|\eta\rangle-|-\eta\rangle)(\langle\eta|-\langle-\eta|)\Big\}_c\\
&~~~~\otimes\Big\{(|\xi\rangle-|-\xi\rangle)(\langle\xi|-\langle-\xi|)\Big\}_d,
\label{ntp-2}
\end{aligned}
\end{equation}
which is not normalized.
The probability of obtaining this result is
\begin{equation}
P_{--}=\frac{(V-1)(V-e^{-\frac{4d^2}{V}})}{2(V^2+e^{-\frac{4d^2}{V}})},
\end{equation}
and it can be simply verified that $P_{++}+P_{--}=1$.
Therefore, only 
the measurement results $|++\rangle$ and $|--\rangle$
can be obtained
in the case of the input state
$\rho^{\Phi(+)}$. This is exactly the same for the case of  $\rho^{\Psi(+)}$. 
In the same way, it can be shown that
if either the input state was $\rho^{\Phi(-)}$ or $\rho^{\Psi(-)}$,
only the measurement results
$|+-\rangle$ and $|-+\rangle$ can be obtained.
 In other words,
the parity of the total incoming state is
perfectly well discriminated by the measurements on single-photon qubits. 

\begin{figure}
\centerline{\scalebox{0.55}{\includegraphics{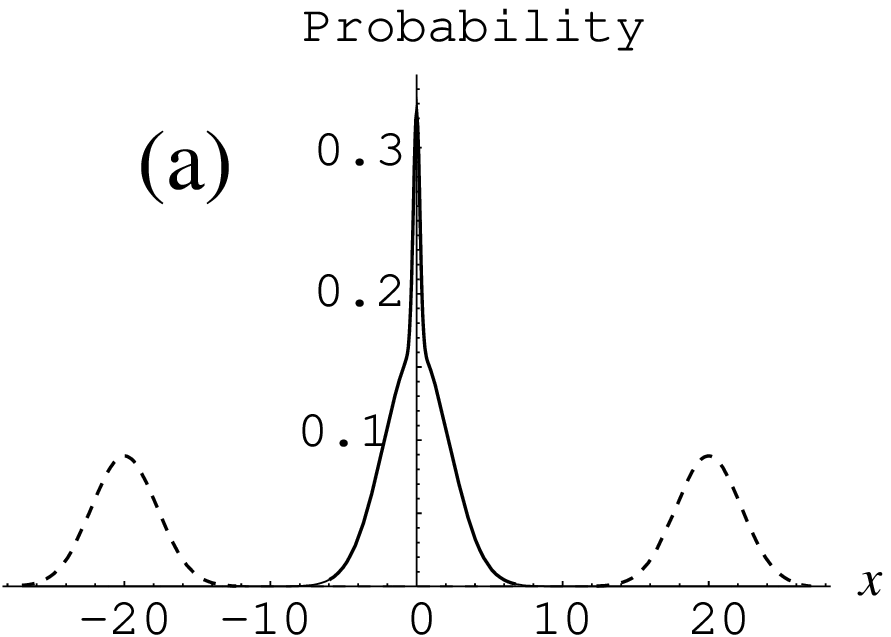}}}
\vspace{0.3cm}
\centerline{\scalebox{0.55}{\includegraphics{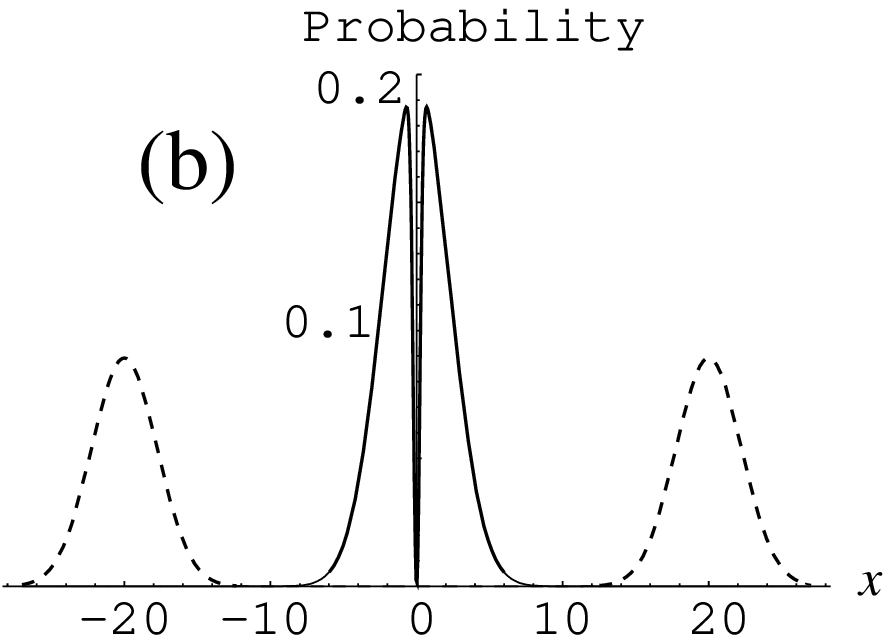}}}
\caption{(a) The probability distributions,
$P^{++}_{\Phi^{(+)}}$ (solid curve) and $P^{++}_{\Psi^{(+)}}$ (dashed curve),
for homodyne measurements at detector C.
(b) The probability distributions,
$P^{++}_{\Phi^{(-)}}$ (solid curve) and $P^{++}_{\Psi^{(-)}}$ (dashed curve),
for homodyne measurements at detector C.
}
\label{fig:P12}
\end{figure}

Subsequently, a homodyne measurement is performed for mode $c$
by homodyne detector C as shown in Fig.~\ref{fig:scheme}(b).
We assume that ideal homodyne measurements are performed, i.e.,
when a homodyne measurement is performed
the state is projected onto
eigenstate $|x\rangle$ of operator $X$
with eigenvalue $x$, where
\begin{equation}
X=\frac{1}{\sqrt{2}}(a +  a^\dagger).
\end{equation}
Let us first consider the case when the measurement result for the 
single photon qubits is $|++\rangle$. In this case, the remaining
state is $\rho^{++}$ in Eq.~(\ref{ntp-1}).
The probability distribution $P^{++}_{\Phi^{(+)}}$
 for the homodyne measurement at detector C
is
\begin{equation}
P^{++}_{\Phi^{(+)}}=\langle x|{\rm Tr}_d[\rho^{++}]|x\rangle
=\frac{V^{\frac{1}{2}}
(e^{-Vx^2}+e^{-\frac{x^2}{V}})
}{\pi^{\frac{1}{2}}(V+1)}. 
\end{equation}
Note that the superscript, $++$,
denotes that the qubit measurement result was 
$|++\rangle$, 
and the subscript, $\Phi^{(+)}$, denotes that 
the input state was $\rho^{\Phi^{(+)}}$. These notations will be used
also for the other cases in this section.
The same analysis can be performed for the other possible measurement outcome
$|--\rangle$:
\begin{equation}
P^{--}_{\Phi^{(+)}}=\langle x|{\rm Tr}_d[\rho^{--}]|x\rangle
=\frac{V^{\frac{1}{2}}
(e^{-Vx^2}-e^{-\frac{x^2}{V}})
}{\pi^{\frac{1}{2}}(V-1)}. 
\end{equation}  
In the same way, for another input state, $\rho^{\Phi(-)}$,
it is straightforward to show:
\begin{equation}
P^{+-}_{\Phi^{(-)}}=P^{++}_{\Phi^{(+)}},~~
P^{-+}_{\Phi^{(-)}}=P^{--}_{\Phi^{(+)}},
\end{equation}
and $P^{++}_{\Phi^{(-)}}=P^{--}_{\Phi^{(-)}}=0$.
On the other hand, 
if the input state was $\rho^{\Psi(+)}$, the probability distributions
$P^{++}_{\Psi^{(+)}}$ and $P^{--}_{\Psi^{(+)}}$ at detector C are
\begin{widetext}
\begin{eqnarray}
&&P^{++}_{\Psi^{(+)}}=\langle x|{\rm Tr}_c[\rho^{++}]|x\rangle
=\frac{V^{\frac{1}{2}}e^{ -\frac{x\{4 d+(2+V^2)x\}}{V}  }
\Big\{
e^{\frac{(1+V^2)x^2}{V}}+2e^\frac{2x(2d+x)}{V}+e^\frac{x(8d+x+V^2 x)}{V}
\Big\}
}{2\pi^{\frac{1}{2}}(e^{\frac{4d^2}{V}}V)+1},\\
&&P^{--}_{\Psi^{(+)}}=\langle x|{\rm Tr}_c[\rho^{--}]|x\rangle 
=\frac{V^{\frac{1}{2}}e^{ -\frac{x\{4 d+(2+V^2)x\}}{V}  }
\Big\{
e^{\frac{(1+V^2)x^2}{V}}-2e^\frac{2x(2d+x)}{V}+e^\frac{x(8d+x+V^2 x)}{V}
\Big\}
}{2\pi^{\frac{1}{2}}(e^{\frac{4d^2}{V}}V)-1}.
\end{eqnarray}
\end{widetext}
It is straightforward to show for the other input state $\rho^{\Psi(-)}$:
\begin{equation}
P^{+-}_{\Psi^{(-)}}=P^{--}_{\Psi^{(+)}},~~
P^{-+}_{\Psi^{(-)}}=P^{++}_{\Psi^{(+)}}.
\end{equation}
The probability distributions $P^{++}_{\Phi^{(\pm)}}$
and $P^{++}_{\Psi^{(\pm)}}$
are plotted in Fig.~\ref{fig:P12}.
Figure~\ref{fig:P12} shows that when the input state was $\rho^{\Phi(+)}$
or $\rho^{\Phi(-)}$, the homodyne measurement outcome by detector C,
characterized by $P^{++}_{\Phi^{(+)}}$ and $P^{--}_{\Phi^{(+)}}$,
is located around the origin. However, when the input state was
$\rho^{\Psi(+)}$ or $\rho^{\Psi(-)}$, the homodyne measurement
outcome by detector C,
characterized by $P^{++}_{\Psi^{(+)}}$ and $P^{--}_{\Psi^{(+)}}$,
is located far from the origin.
Therefore, two of the Bell states, $\rho^{\Phi(+)}$ or $\rho^{\Phi(-)}$,
can be well distinguished from the other two by the homodyne
detector C for the case of the measurement outcome $|++\rangle$.
Finally,
by combining the homodyne measurement result and the qubit measurement result,
all four Bell states can be effectively distinguished.
For example, let us assume that the measurement outcome of the 
single photon detectors was $|++\rangle$ and the homodyne detection outcome was
around the origin, i.e., $x\approx0$.
Then, one can say that state $\rho^{\Psi(-)}$ has been measured for the result
of the thermal-Bell measurement.

\begin{figure}
\centerline{\scalebox{0.55}{\includegraphics{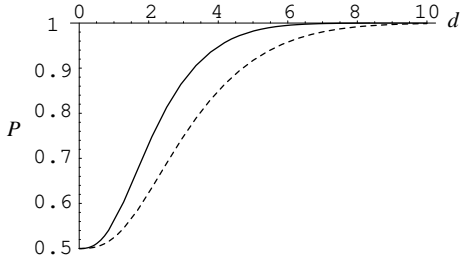}}}
\caption{The distinguishability $P_s$ between states $\rho^{\Psi(+)}$ and $\rho^{\Phi(+)}$
by a homodyne measurement against for $V=10$ (solid curve) and $V=20$ (dashed curve)
against distance $d$. See text for details.}
\label{fig:Ps}
\end{figure}

As implied in Fig.~\ref{fig:P12},
 the overlaps between the probability distributions
around the origin, $P^{++}_{\Phi^{(+)}}$ and $P^{--}_{\Phi^{(+)}}$,
and the other distributions,
$P^{++}_{\Psi^{(+)}}$ and $P^{--}_{\Psi^{(+)}}$,
are extremely small for a sufficiently large $d$.
In other words, the distinguishability
 by the homodyne detection
rapidly approaches 1 as $d$ increases.
As an example, we can calculate the distinguishability between
the states $\rho^{\Psi(+)}$ and $\rho^{\Phi(+)}$ by the homodyne measurement
by detector C.
The distinguishability  by homodyne detection is 
\begin{equation}
{\cal P}_s= \frac{1}{2}\Big\{ \int_{|x|<d}dx P^{++}_c(x)
+\int_{|x|\geq d} dx P^{++}_d(x)
\Big\}
\end{equation}
which is plotted in Fig.~\ref{fig:Ps}.
The distinguishability is 
 ${\cal P}_s \approx 0.99$ for $d=5.5$ ($d=7.8$)
 when $V=10$ ($V=20$), and it
becomes as high as 
 ${\cal P}_s> 0.99999$ for $d=10$ ($d=15$) when $V=10$ ($V=20$). 
If necessary, another homodyne measurement can be performed
for mode $d$ to enhance distinguishability of the Bell measurement.
When the probability distribution at detector C is around the origin
that of detector D is far from the origin and vice versa.

Note also that the second scheme using homodyne detection
is robust to detection inefficiency compared with the first scheme 
using photon number resolving measurements.
In the first scheme, 
even if a detector misses only one photon, it will result in
a completely wrong measurement outcome.
In the second scheme, however, the measurement outcome will not
be affected in that way.
If a single photon detector misses a photon, it will be immediately
recognized. Such a case can simply be discarded so that
it will only degrade the success probability of the Bell measurement.
The homodyne detection inefficiency 
will not significantly affect the result when
the distributions around the origin
and the distributions far from the origin 
are well separated, i.e., when $d\gg \sqrt{V}$,
as shown in Fig.~\ref{fig:P12}.
On the other hand, loss in the Kerr medium will have
a detrimental affect.

\subsection{Quantum teleportation and computation}

Quantum teleportation of a thermal-state qubit
can be performed using one of the Bell states
as the quantum channel.
Let us assume that
Alice needs to teleport a thermal-state qubit, $\rho^{\psi}$,
to Bob using a thermal-state entanglement,  $\rho^{\Psi(-)}$,
shared by the two parties.
The total state can be represented as
\begin{widetext} 
\begin{equation}
\rho^{\psi}_{1}\otimes\rho^{\Psi(-)}_{23}=N_t
\int d\alpha^2 d\beta^2 d\gamma^2
P^{th}_\alpha(V,d)
P^{th}_\beta(V,d)
P^{th}_\gamma(V,d)
\Big[(a|\alpha\rangle+b|-\alpha\rangle)_1
(|\beta,-\gamma\rangle-|-\beta,\gamma\rangle)_{23}\Big]
\Big[h.c.\Big].
\end{equation}
\end{widetext}
Alice first needs to perform the
thermal-Bell measurement described in
the previous subsection. To complete the teleportation process,
Bob should perform an appropriate
unitary transformation on his part of the quantum channel
according to the measurement result sent from Alice
via a classical channel. It is straightforward to show
that the required transformations are
exactly the same to those for the coherent-state qubit \cite{JKL01}.
When the measurement outcome is $\rho^{\Psi(-)}$, 
Bob obtains a perfect replica of the 
original unknown qubit without any operation.
When the measurement outcome is $\rho^{\Phi(-)}$, Bob should perform
$|\alpha\rangle\leftrightarrow|-\alpha\rangle$ on his qubit
in Eq.~(\ref{t-qubit2}). Such a phase
shift by $\pi$ can be done using a phase shifter whose action is
described by $P(\varphi)=\mbox{e}^{i\varphi a^\dag a}$,
where $a$ and $a^\dagger$ are the annihilation and creation operators.
When the outcome is $ \rho^{\Psi(+)}$, 
the transformation should be performed as 
$|\alpha\rangle\rightarrow|\alpha\rangle$ and
$|-\alpha\rangle\rightarrow-|-\alpha\rangle$.
It is known that the displacement operator is
a good approximation of this transformation for $d\gg1$ \cite{Jeong02}.
This transformation can also be achieved 
by teleporting the state again locally and
repeating until the required phase shift is obtained \cite{Ralph03}.
When the outcome is $\rho^{\Phi(+)}$, $\sigma_x$ and
$\sigma_z$ should be successively applied.

\section{Conclusion}

In this paper, 
we have studied characteristics of 
superpositions and entanglement of
thermal states at high temperatures
and discussed their applications to quantum information processing.
The superpositions and entanglement of
thermal states show various nonclassical properties such as
interference patterns, negativity of the Wigner functions, and
violations of the Bell-CHSH inequality.   
The Bell violations
are more sensitive to the interaction time during the generation process  
when the thermal temperature
(i.e. mixedness) of the thermal-state entanglement is larger.
Therefore, in order to observe the Bell violations
using the mixed state at a high temperature,
the interaction time in the Kerr medium
should be accurate.
We have pointed
out that certain superpositions of high-temperature thermal states,
symmetric in the phase space, can also be
 generated. 
Some of these states have neither squeezing properties nor
negative values in their Wigner functions but
they are found to be highly nonclassical.

We have introduced the thermal-state qubit and thermal-Bell states
for applications to quantum information processing.
We have presented two possible methods for
the Bell-state measurement.
The Bell-state measurement enables one to perform
quantum teleportation and gate operations 
for quantum computation with thermal-state qubits.
The first scheme uses two photon number resolving
detectors and a 50-50 beam splitter
to discriminate the thermal-Bell states.
Using the second scheme, 
it is possible to effectively discriminate
the thermal-Bell states
without photon number resolving
detection. The required resources for the second scheme
are two Kerr nonlinear interactions,
two single photon detectors, two 50:50 beam splitters and one homodyne detector. 
The second scheme is more robust to inefficiency of the detectors:
the inefficiency of the single photon detectors
only degrades the success probability of the Bell measurement.

\acknowledgments

This work was supported by 
 the DTO-funded U.S. Army Research Office Contract
No. W911NF-05-0397, the Australian Research Council and
Queensland State Government.


\begin{references}
\bibitem{ref19} We note that denominator $V$ was missing in the
generation probability ${\cal P_\pm}$ in~\cite{jr06}.
\end{references}
\end{document}